\documentclass[twocolumn]{aastex61}
\pdfoutput=1 %for arXiv submission
\usepackage{amsmath,amstext}
\usepackage[T1]{fontenc}
\usepackage{apjfonts} 
\usepackage[figure,figure*]{hypcap}

\newcommand{\gps}{\ensuremath{g_{\rm P1}}}
\newcommand{\rps}{\ensuremath{r_{\rm P1}}}
\newcommand{\ips}{\ensuremath{i_{\rm P1}}}
\newcommand{\zps}{\ensuremath{z_{\rm P1}}}
\newcommand{\yps}{\ensuremath{y_{\rm P1}}}
\newcommand{\wps}{\ensuremath{w_{\rm P1}}}
\newcommand{\grizy}{g_{\rm P1}\rps\ips\zps\yps}

\newcommand{\PS}{\protect \hbox {Pan-STARRS}}

\shorttitle{Shape Distribution of NEOs}
\shortauthors{McNeill et al.}

\begin{document}

\title{Constraining the Shape Distribution of Near Earth Objects from Partial Lightcurves}

\author{A. McNeill}
\affiliation{Department of Physics and Astronomy, Northern Arizona University, Flagstaff, AZ 86011, USA}
\author{J.L. Hora}
\affiliation{Center for Astrophysics | Harvard \& Smithsonian, 60 Garden St., MS-65, Cambridge, MA 02138-1516, USA}
\author{A. Gustafsson}
\affiliation{Department of Physics and Astronomy, Northern Arizona University, Flagstaff, AZ 86011, USA}
\author{D.E. Trilling}
\affiliation{Department of Physics and Astronomy, Northern Arizona University, Flagstaff, AZ 86011, USA}
\author{M. Mommert}
\affiliation{Lowell Observatory, 1400 W.\ Mars Hill Rd., Flagstaff, AZ, 86001, USA}

\begin{abstract}
In the absence of dense photometry for a large population of Near Earth Objects (NEOs), the best method of obtaining a shape distribution comes from sparse photometry and partial lightcurves. We have used 867 partial lightcurves obtained by Spitzer to determine a shape distribution for sub-kilometre NEOs. From this data we find a best fit average elongation $\frac{b}{a}=0.72 \pm 0.08$. We compare this result with a shape distribution obtained from 1869 NEOs in the same size range observed by Pan-STARRS 1 and find the Spitzer-obtained elongation to be in excellent agreement with this PS1 value of $\frac{b}{a}=0.70 \pm 0.10$. These values are also in agreement with literature values for $1<D<10$ km objects in the main asteroid belt, however, there is a size discrepancy between the two datasets. Using a smaller sample of NEOs in the size range $1<D<5$ km from PS1 data, we obtain an average axis ratio $b/a = 0.70 \pm 0.12$. This is more elongated than the shape distribution for main belt objects in the same size regime, although the current uncertainties are sizeable and this should be verified using a larger data set. As future large surveys come online it will be possible to observe smaller main belt asteroids to allow for better comparisons of different sub-kilometre populations. 

\end{abstract}


\section{Introduction}

Near-Earth Objects (NEOs) are defined as minor planets with perihelion distance q < 1.3 AU.  The majority of NEOs are either Apollos or Atens both of which have Earth-crossing orbits, with Apollos spending most of their orbit outside that of the Earth and Atens spending most of their orbit inside that of the Earth.

The shape and rotational properties of NEOs are governed by the interplay of several different mechanisms. These include collisional effects and thermal forces such as the Yarkovsky and YORP effects. The primary method of identifying the shape and spin information of asteroids is through lightcurve inversion. Of the 18748 known NEOs, as of September 2018, only 1252 have reliable lightcurves recorded in the Lightcurve Database (LCDB; \citealt{warner2009}, Last Updated 23 June 2018). With multiple epochs of observation it is possible to obtain a well-constrained model of the convex approximation of an asteroid's shape as well as its spin pole axis (\citealt{kaas2001}; \citealt{durech2016}). To date around 1000 asteroid shape models have been obtained and are stored in the Database of Asteroid Models from Inversion Techniques (DAMIT; \citealt{durech2010}). Of these, 59 are NEOs. This is clearly not enough to determine a shape distribution for the NEO population.

In lieu of a large sample of shape models, sparse photometric data and partial lightcurves can be used to determine an estimate for the shape distribution of a population. Previous work on this has been carried out in \cite{mcneill2016} and \cite{cibulkova2017}. Both used sparse photometry from the Pan-STARRS 1 Survey to obtain a shape distribution for main belt asteroids (MBAs). Both studies were in good agreement that, if the objects are assumed to be a population of prolate spheroids ($a > b \geq c$), the average axis ratio for a small ($1 < D < 10$ km) main belt object is approximately $\frac{b}{a}=0.8$.

Obtaining a shape distribution in this way is a more difficult proposition for NEOs due to their lesser numbers and the greater influence of phase angle effects. A preliminary version of an investigation of the shape distribution of NEOs was carried out by \cite{mcneill_2017} and an updated form is presented in this paper. A comparison of the shape distribution of NEOs and MBAs may offer insight into the process by which the main belt "resupplies" NEO space. At present this is non-trivial as the size ranges accessible in both populations are not the same, with a dearth of data available for small ($D < 1$ km) MBAs and a relatively low abundance of large kilometre-sized NEOs. This will be particularly interesting in the future when larger surveys such as the Large Synoptic Survey Telescope \citep[LSST;][]{veres2017} come online allowing for a more complete study of smaller MBAs than ever before.

Here we present the shape distributions determined from sub-kilometre NEO populations observed by both the Spitzer Space Telescope and Pan-STARRS 1. We also find a shape distribution for larger NEOs from Pan-STARRS 1 in the size range $1<D<5$ km. In Section 6 we compare and contrast these values with the shape distribution for MBAs from literature.

\section{Observations}

\subsection{Spitzer Space Telescope}

The observations used in this project were obtained with the {\it Spitzer}/IRAC instrument \citep{fazio2004} as part of the NEO Survey (Program ID 11002), and the NEO Legacy Survey (Program ID 13006) [\citealt{trilling2010}, \citealt{trilling2016}]. The observations were carried out using the moving single mode to track the NEO motion, dithering during the observations to remove the effect of bad pixels or heterogenous sensitivity across the array. The observing window for each object was typically chosen to be the period when the NEO would have close to its maximum flux as seen by {\it Spitzer}, in order to minimize the time necessary to detect the source. We observed only those objects with small positional uncertainties (<150$''$) as seen by {\it Spitzer} to ensure that the targets would fall within the IRAC field of view. 

The exposure time for the observations was determined to allow detection of the source at a $10\sigma$ level, using an estimate of the NEO flux density as observed from {\it Spitzer} based on the Horizons {\it H} magnitude. A minimum apparent motion of the source relative to the background during the observation was also applied, to make it possible to distinguish the NEO from background sources. For slow moving objects the number of frames was increased or a second epoch of observations was carried out to allow background subtraction and photometry of the object to be carried out successfully. 

The observations were made using the 4.5~$\mu$m band where the NEO is usually much brighter, which reduced the required integration time as compared to the 3.6~$\mu$m band and allowed us to maximize the number of objects observed in the time allocated. The minimum observation period was set to approximately 30 minutes in order to ensure sufficient frames were created to allow background subtraction and object identification. The maximum time for objects in the survey was chosen to be 3 hours, to keep the total time requested within the range allowed by the {\it Spitzer} program. For further information on this and the data reduction processes see \citet{trilling2016} and \citet{hora2018}.

\subsection{Pan-STARRS 1}

The \PS 1 1.8m telescope on Haleakala contains a 1.4 gigapixel orthogonal transfer array CCD camera (GPC1) and covers a field of view of $\sim 7$ square degrees on the sky (\citealt{denneau2013}). GPC1 is made up of an $8 \times 8$ grid of OTA CCDs with each of these OTAs in turn comprising of an $8\times8$ array of $590\times598$ $10\mu m$ CCDs (\citealt{tonry2012}). The exposure time of observations is survey/filter dependent and the CCD readout time is approximately $7$ seconds. The system operates by taking a 45 second exposure of an area of sky and then returning to the same area after approximately 15 minutes. These images are detrended and calibrated via the Image Processing Pipeline (IPP; \citealt{Magnier2013}). The IPP also subtracts consecutive pairs of images, detects objects remaining in these difference images and passes those detections to the Moving Object Processing System (MOPS; \citealt{denneau2013}). MOPS attempts to link detections of transient objects into `tracklets' containing the same moving objects, and associate them with previously discovered Solar system bodies.

The $\grizy$$\wps$ filter system used by the survey is similar to the Sloan-Gunn system, but there are slight differences in wavelength range (\citealt{tonry2012}). In this work we deal exclusively with data taken using the $\wps$-band filter. The $\wps$ filter spans the combined range of the $\gps$, $\rps$ and $\ips$ filters allowing detections down to 22nd magnitude.

During the first 18 months of survey time, PS1 made approximately 1.5 million confirmed detections of moving objects, to which we applied a series of constraints. Only detections with a formal magnitude uncertainty of $\leq 0.02$ and at a phase angle $\leq 20^{\circ}$ were considered. This should ensure that magnitudes were not significantly affected by photometric uncertainties. This survey data was part of the first public data release in late 2016 (\citealt{magnier2016}).

\section{Partial Lightcurves from Spitzer}

\begin{figure}
  \begin{center}
\includegraphics[width=0.45\textwidth]{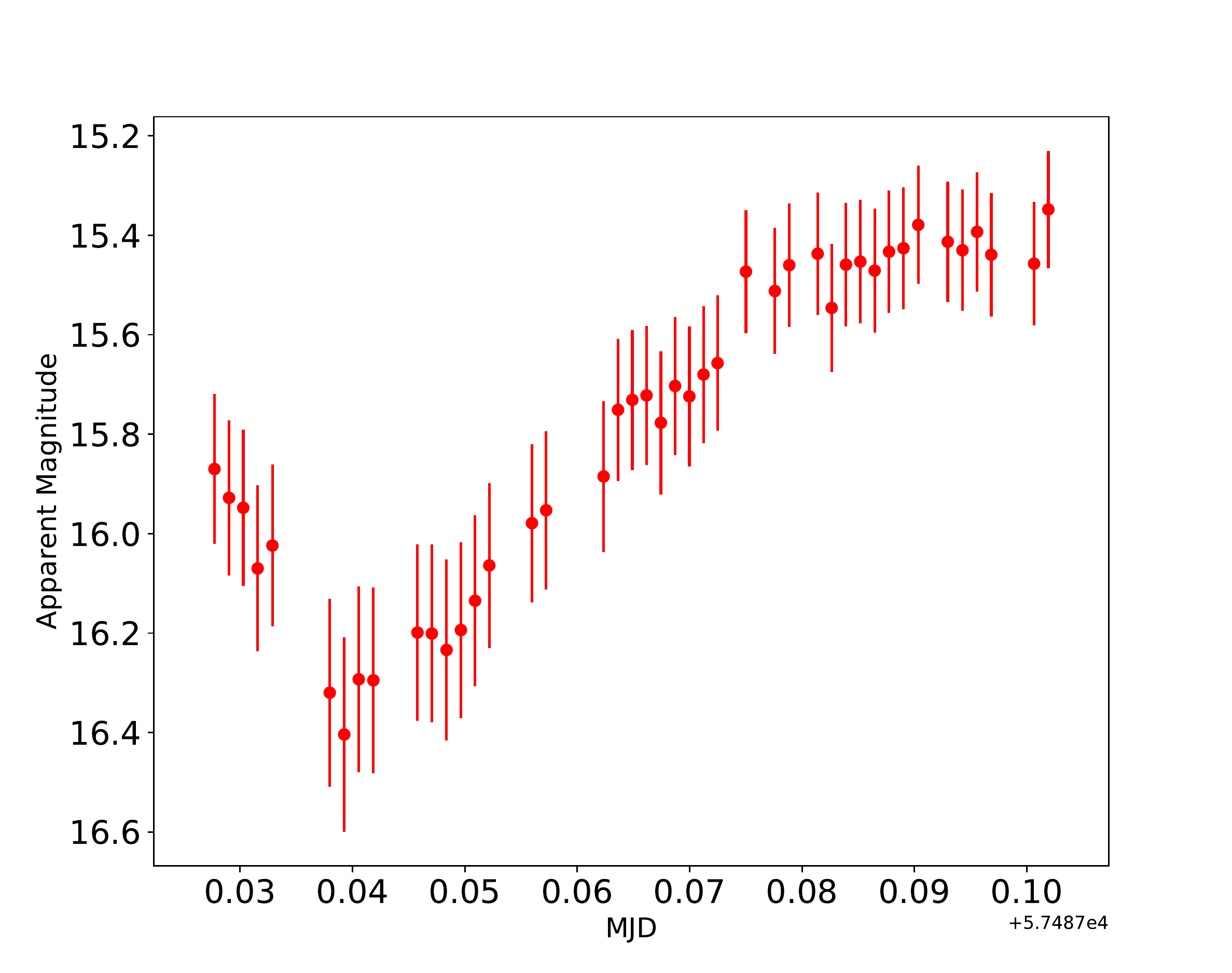}
\includegraphics[width=0.45\textwidth]{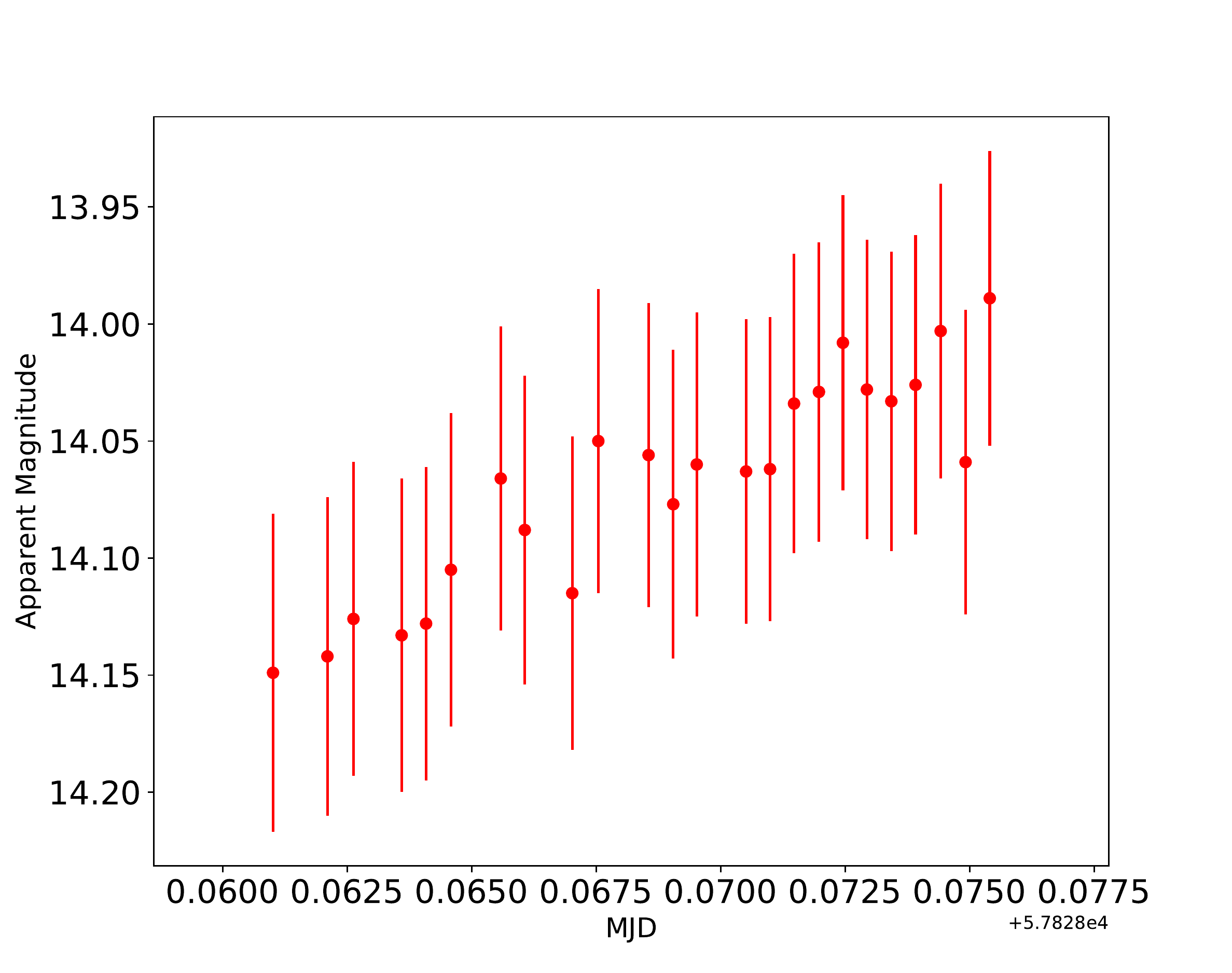}   
\includegraphics[width=0.45\textwidth]{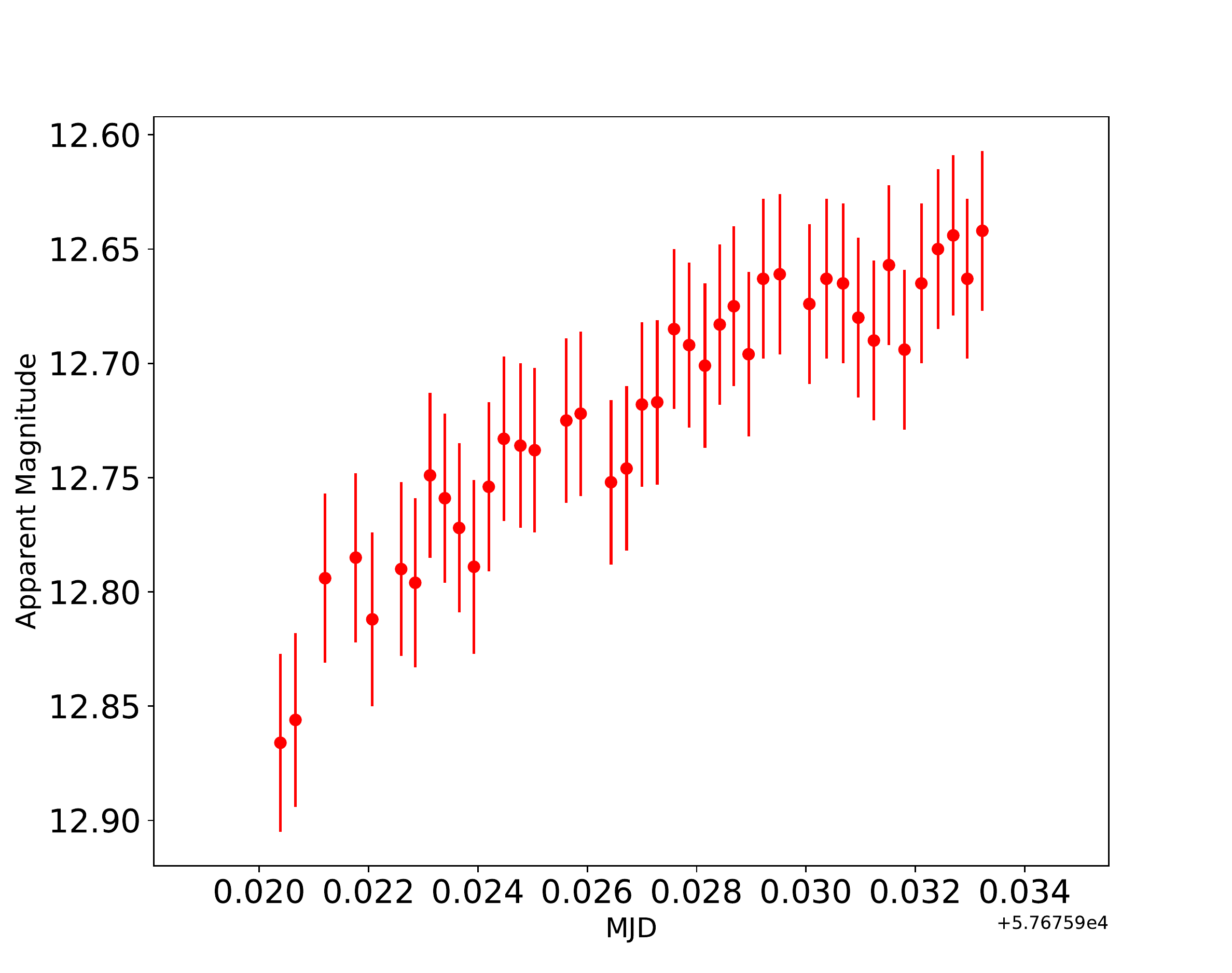}   
\caption{Partial lightcurves for 2001 SA270, 2001 AU47 and 5025 P-L, three of the 867 NEOs observed by {\it Spitzer} considered in this work.}
\label{lc}
  \end{center}
\end{figure}

\begin{figure}
  \begin{center}
\includegraphics[width=0.45\textwidth]{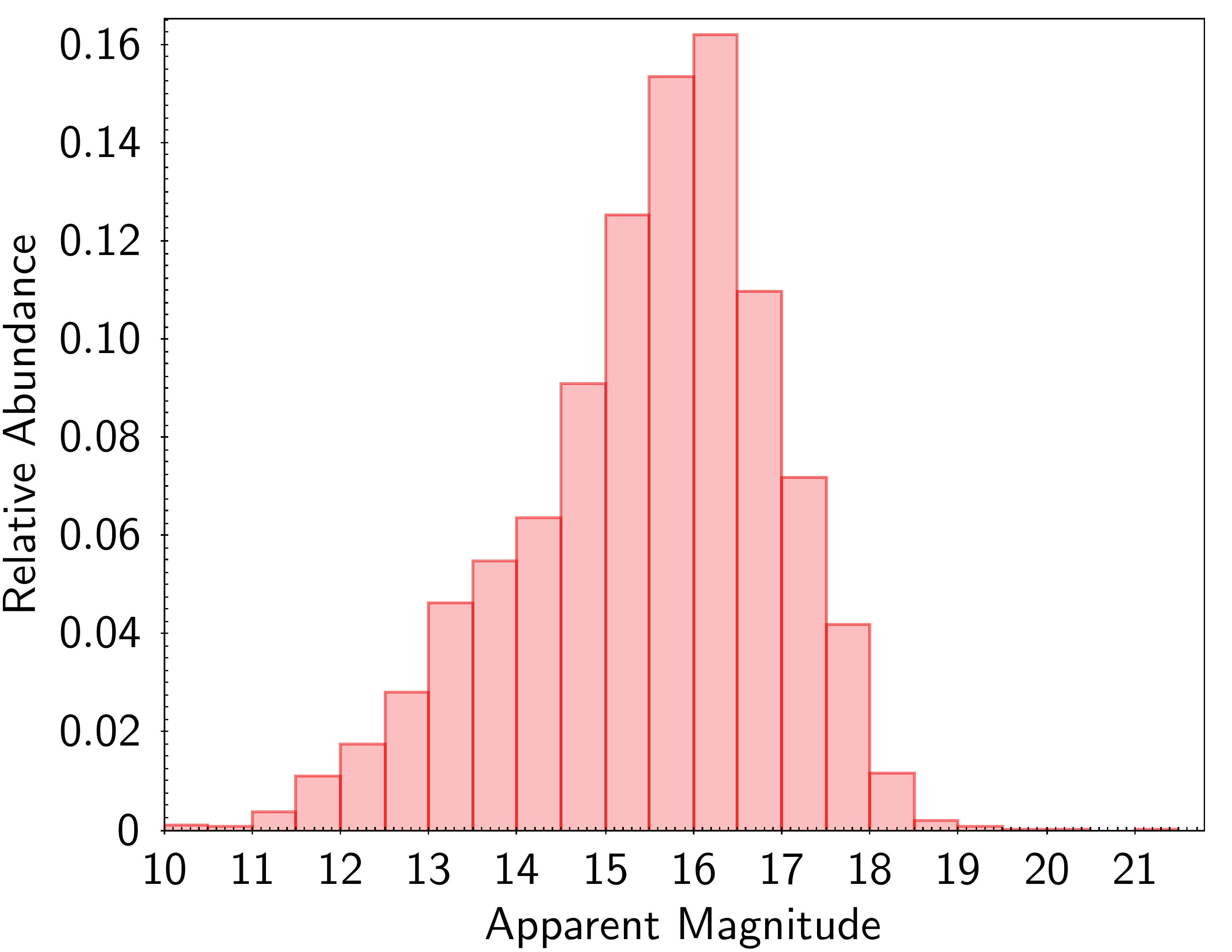}   
\caption{Histogram showing the distribution of apparent magnitudes from the Spitzer observations used in this investigation.}
\label{fig:spitzmag}
  \end{center}
\end{figure}

Across both the NEOLegacy and NEOSurvey datasets we have 867 partial lightcurves for sub-kilometre Near Earth Objects, once a minimum signal-to-noise threshold is used. The timespan for these lightcurves range from several minutes up to 3 hours. The average time between two consecutive data points in these curves is 2.5 minutes with the partial light curves studied here having between 5-90 data points. To search for periodicity in the partial lightcurves we utilise the Generalised Lomb-Scargle Periodogram (\citealt{zechmeister2009}). As the majority of the partial lightcurves are fragmentary, it is impossible to obtain accurate period values. In most cases, only a lower limit on rotation period can be established at best. Example partial lightcurves are presented in Figure~\ref{lc}. We identified no objects in this sample to have rapid rotation periods. The distribution of apparent magnitudes from Spitzer observations is given in Figure~\ref{fig:spitzmag}.

\begin{figure}
  \begin{center}
\includegraphics[width=0.45\textwidth]{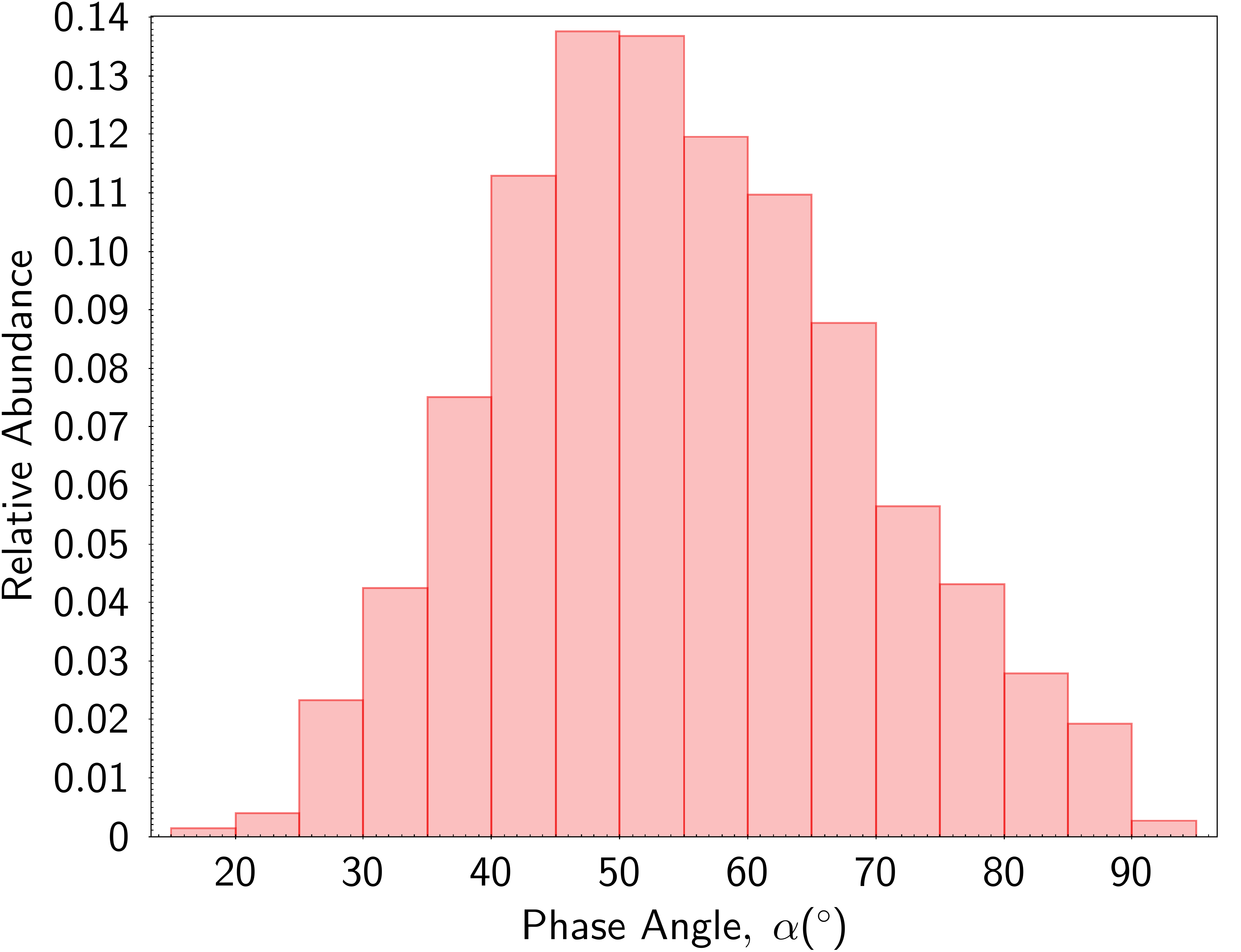}   
\caption{Histogram showing the distribution of phase angles from the {\it Spitzer} observations used in this investigation.}
\label{alphahist}
  \end{center}
\end{figure}

Similarly, the only direct estimate of shape that can be made from each individual asteroid is a lower limit on the body's elongation based upon the amplitude of the partial lightcurves as corrected accounting for the phase angle of the observations. A histogram showing the distribution of phase angles from the {\it Spitzer} observations used in this investigation is given in Figure~\ref{alphahist}. The relatively large phase angles of these observations means that we have to ensure that phase angle effects are properly accounted for.

\cite{zappala1990} analysed the amplitude-phase relationship for asteroids. This relationship is shown to be linear for phase angles, $\alpha < 40^{\circ}$. At these values the effect of the phase angle on the estimated lower limit elongation of the object can be expressed as Equation~\ref{eqn:alpha} where $s$ is some slope parameter dependent on composition (\citealt{thirouin2016}). For this work we assume a conservative slope parameter of $s = 0.015$ mag deg$^{-1}$.

\begin{equation}
\frac{b}{a} \geq 10^{-\frac{0.4\Delta m}{1+s\alpha}}
\label{eqn:alpha}
\end{equation}

The majority of our observations with {\it Spitzer} are made at much greater phase angles where this approximation does not hold. Instead, we assume the non-linear relationship from Figure~1 of \cite{zappala1990} as approximated with a polynomial fit.

In order to obtain a reliable shape and spin pole model for the individual asteroids, observations at multiple epochs and observing geometries would be required. However, none of the objects in this sample had sufficient repeat observations to allow for this.  

\section{Partial Lightcurves from Pan-STARRS 1}

From the first 18 months of the Pan-STARRS 1 survey we have 9442 individual detections of 1869 sub-kilometre Near Earth Objects. For the vast majority of these objects we have 4-8 detections. Consecutive detections from Pan-STARRS are associated into 'tracklets' of 4 data points over the space of approximately one hour. More detail on the data selection methods and statistical information regarding the PS1 dataset can be found in \cite{mcneill2016} and \cite{mcneill_2017}.

\begin{figure}
  \begin{center}
\includegraphics[width=0.45\textwidth]{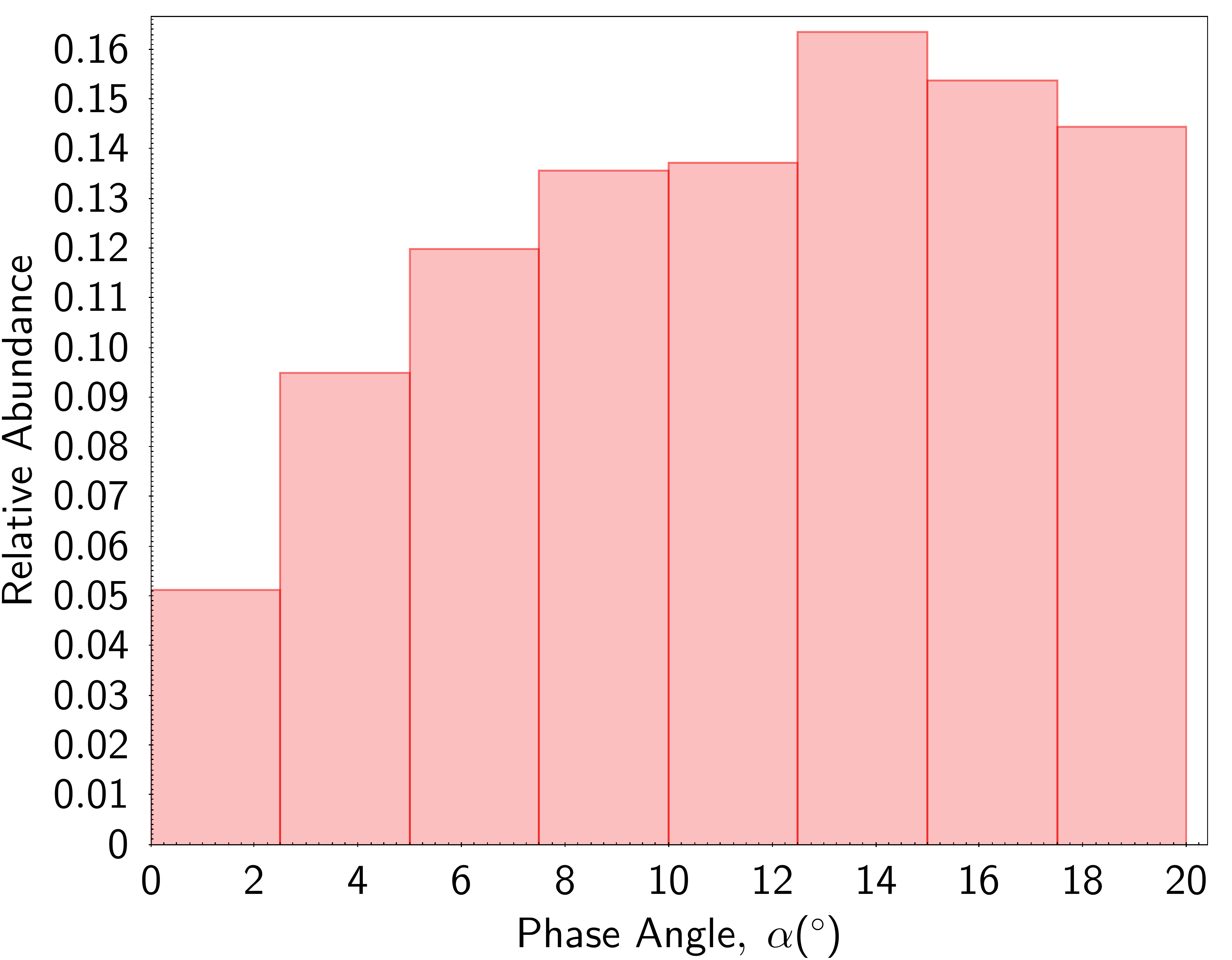}   
\caption{Histogram showing the distribution of phase angles from the Pan-STARRS observations used in this investigation.}
\label{fig:ps1_alpha}
  \end{center}
\end{figure}

The phase angle distribution of these detections is given in Figure~\ref{fig:ps1_alpha}. The range of phase angles considered here is in the range at which the phase-angle amplitude effect acts linearly and hence we make use of Equation 1 to correct the measured amplitudes. Using a cut-off in photometric uncertainty of $< 0.02$ magnitudes we have a sample of 1570 detections of approximately 350 NEOs in the size range $1<D<5$ km. 

\section{Shape Distribution for NEOs}

To constrain the shape distribution of our sample of NEOs we use a more advanced version of the model described in \cite{mcneill2016} which can more accurately account for effects of orbital geometry e.g. phase-angle amplitude effects. The model assumes a population of triaxial ellipsoids with axes $a > b \geq c$ generated from an input distribution of $b/a$ axis ratios and spin pole axes. For each object, the projected area at some point about the object's rotation is determined from Equation~\ref{eqn:area} where $A$ is the apparent cross-section of the asteroid with principal axes, $a$, $b$ and $c$ as seen from Earth, 
$\theta$ is the angle between the spin axis of the asteroid and the plane of the sky, and $\phi$ is the rotational phase. This model has been thoroughly tested using synthetic populations of known shape distribution and correctly reproduces the best fit in each case for samples $n > 300$, for more detail see \cite{mcneill_2017}.

\begin{equation}
A=\pi \sqrt{c^{2}sin^{2}\theta(a^{2}sin^{2}\phi + b^{2}cos^{2}\phi)+a^{2}b^{2}cos^{2}\theta}
\label{eqn:area}
\end{equation}

The synthetic population is sampled using the same cadence as the {\it Spitzer} and PS1 observations, respectively, and for total epochs consistent with the partial lightcurves. The model produces a synthetic population of the same size as the input sample to allow for two-sampled chi-squared testing to be carried out. This is repeated many times for each population. The model also assumes an uncertainty on each measurement derived from the distribution of uncertainties in the measured magnitudes from the observed data. The magnitudes generated by the model are assumed entirely to be due to geometric effects (i.e., albedo variegation effects and limb darkening are not accounted for). To quantitatively determine the goodness of the fit for the amplitude distribution of each model population, we employ a simple two-sampled chi-squared test and obtain a best fit by minimizing this value. Assuming that all asteroids are prolate spheroids ($a > b = c$) the value of $b/a$ is varied as a truncated Gaussian with a centre $\mu$ and standard deviation $\sigma_{G}$. The Gaussian distribution is truncated at the point where $b/a = 1$. We assume the spin pole distribution for NEOs determined by \cite{tardioli2017}. The model uses a uniform spin frequency distribution
from $1-10.9$ day$^{-1}$ across all applicable size ranges, corresponding
to rotational periods from that of the spin barrier at
$2.2$ h to a period of $24$ h. This assumption is reasonable when
compared with the flat distribution of measured rotational
frequencies at small sizes and with the relative scarcity of extremely fast or extremely slow rotators (\citealt{pravec2002}). We also used an iteration of the model assigning rotation periods to model objects based on the distribution of rotation periods for known NEOs from the Light Curve Database (\citealt{warner2009}). In both cases statistically identical answers were produced. We assume all fits for which $\chi^{2} \leq 2\chi^{2}_{min}$ to be equally valid and set the uncertainties on our best fit to be described by the range of these values.

\begin{figure}
  \begin{center}
\includegraphics[width=0.45\textwidth]{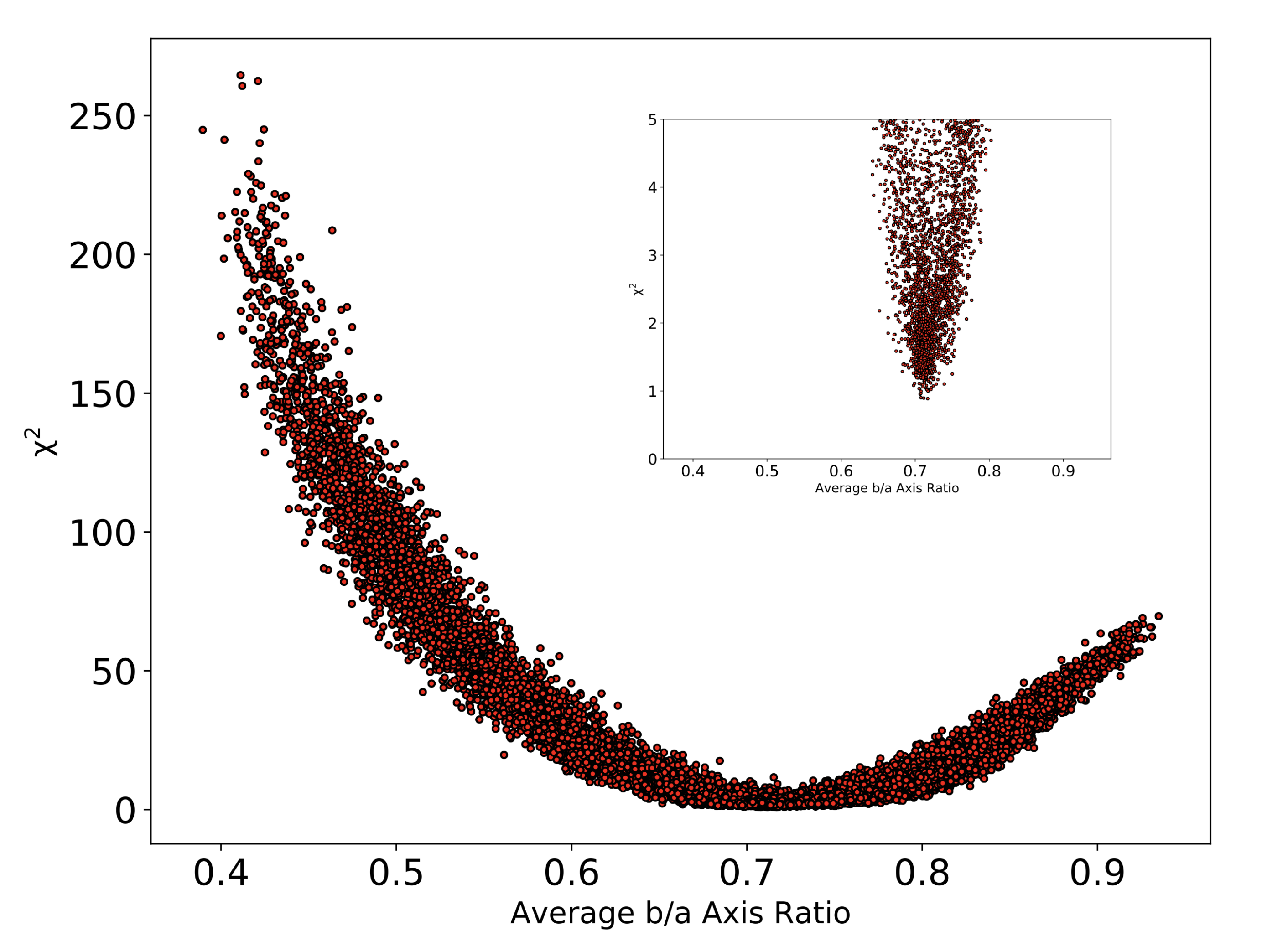}   
\caption{Chi-squared statistics from model asteroid populations compared to partial NEO lightcurves from Spitzer as a function of axis ratio. Each point shows the chi-squared value determined by comparing the observed lightcurves to those of a synthetic population with an assumed Gaussian shape distribution. The axis ratio value corresponds to the mean shape for each model population. The inset plot shows a zoomed-in view of the chi-squared minimum}.
\label{spitzerchi}
  \end{center}
 \end{figure}

Assuming a population of prolate spheroids ($a\geq b = c$) we find the best fit average axis ratio to be $b/a = 0.72 \pm 0.08$. This corresponds to a truncated Gaussian with centre $\mu = 0.79 \pm 0.07$ and standard deviation $\sigma_{G}=0.22 \pm 0.04$. A plot of chi-squared against $b/a$ is given in Figure~\ref{spitzerchi}. This value is in excellent agreement with the average shape of objects determined from collisional lab experiments suggesting a collisional formation mechanism for these objects (\citealt{michikami2016}).

\begin{figure}
  \begin{center}
\includegraphics[width=0.45\textwidth]{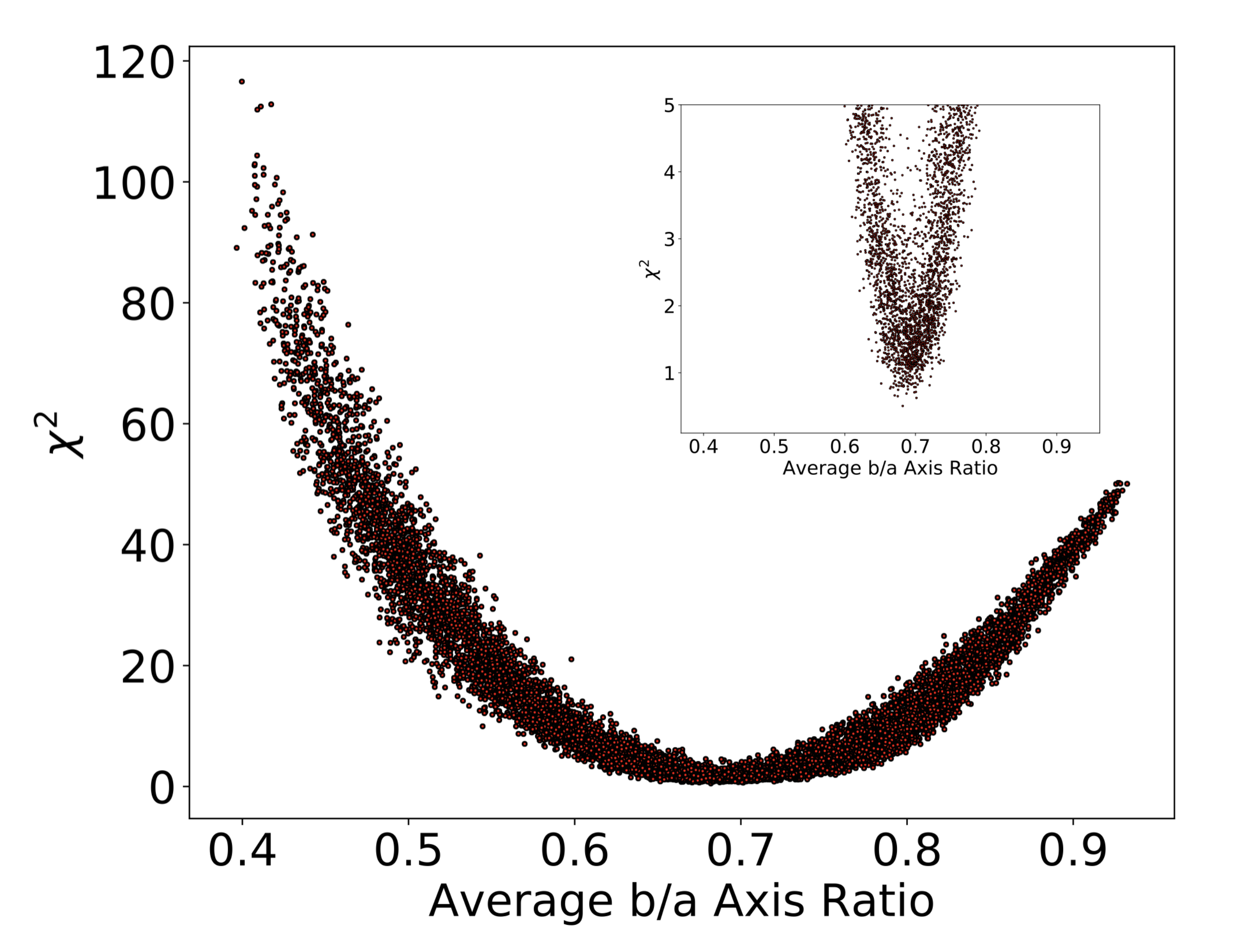}   
\caption{The same plot as Figure~\ref{spitzerchi} except using the NEO detections from Pan-STARRS 1.}
\label{ps1chi}
  \end{center}
 \end{figure}

We also used Pan-STARRS 1 detections for NEOs to generate a shape distribution for NEOs in the same size range as the {\it Spitzer} dataset. We utilised approximately 9000 detections of 1869 unique NEOs from the initial 18 months of the PS1 survey and find a best fit corresponding to an average axis ratio $b/a = 0.70 \pm 0.10$ using the same criteria as the {\it Spitzer} data in order to obtain the uncertainty in this value. This corresponds to a truncated Gaussian with centre $\mu = 0.76 \pm 0.08$ and standard deviation $\sigma_{G}=0.22 \pm 0.05$ A plot showing the chi-squared values obtained for a range of models is given in Figure~\ref{ps1chi}. 

To ensure that the result we obtain is not a result of the photometric cut-offs applied to our data set this process was repeated for data sets recreated using only data with lower photometric uncertainty. This did not significantly alter the result of the model, the only notable change was the slightly larger uncertainty on the obtained value due to the smaller sample size.

\section{Discussion}
\label{sec:discussion}

The determined average values for the NEO populations agree to within uncertainties. We also compare these shape distribution results to the most recent distributions obtained from main belt asteroids. \cite{cibulkova2017} derived a shape distribution for MBAs using survey data from Pan-STARRS 1. They report an average axis ratio for these objects of $b/a = 0.8$ which agrees to within uncertainties of the NEO result presented in this paper. \cite{erasmus2018} derived an average shape for MBAs using 1000 serendipitously obtained MBA partial lightcurves and found an average axis ratio for their sample of $b/a = 0.74 \pm 0.07$, also in agreement with our result. It is worth noting that the size range for the MBAs in previous studies and that for NEOs in this work are not the same. The MBA data in each of the two previously mentioned works falls in the size range $1 < D < 10$ km, as data samples for these objects are not complete at smaller diameters. When LSST and other large surveys come online, it will be possible to obtain lightcurve data for much smaller MBAs allowing for a more appropriate comparison.\\

We also constrain a shape distribution for kilometre-size NEOs for comparison with previous studies of main belt shape distribution. For objects in the size range $1<D<5$ km we find an average axis ratio $b/a = 0.70 \pm 0.12$, in agreement with the value obtained for sub-kilometre NEOs in this work. We see that this value is more elongated than the value for main belt asteroids of the same size, although the sizeable uncertainties involved do overlap. In order to clearly show the difference in elongation between these two populations at similar diameter, a larger data-set will be required. The results to date suggest a trend toward more elongation in NEOs compared to MBAs, although at present the values are the same within uncertainties. Although this trend may not yet be statistically significant we consider what its explanation may be as with larger data sets in future the uncertainties on these values will decrease and this may become significant.

\begin{figure}
  \begin{center}
\includegraphics[width=0.5\textwidth]{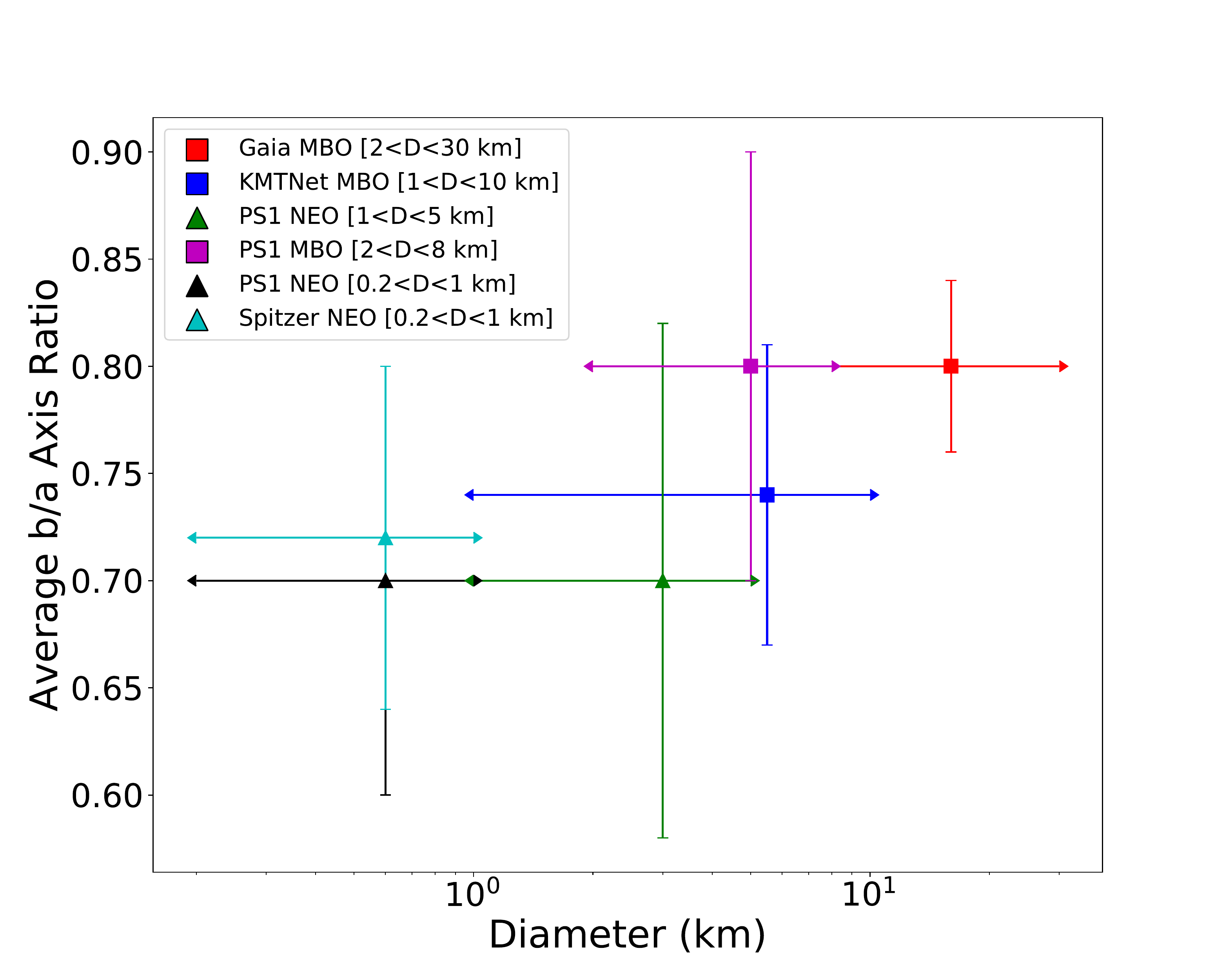}   
\caption{A plot showing the average axis ratios determined for different asteroid populations and size ranges. The arrow ranges in the x-axis show the diameter range contained within each data set. Here in all cases we assume that $a>b=c$. References: \cite{mommert2018}, \cite{erasmus2018}, \cite{mcneill_2017}, \cite{cibulkova2017} and this paper, respectively according to their position in the figure legend.}
\label{fig:shapesumm}
  \end{center}
 \end{figure}

Figure~\ref{fig:shapesumm} shows the average axis ratios obtained for NEOs and MBAs from this work and previous studies as a function of diameter. We see that NEOs are more elongated than main belt objects of the same size in the range $1<D<5$ km. We also see that currently there is no shape distribution obtained for sub-kilometre MBAs for comparison to the NEO distribution. To obtain this a larger, more complete, data-set for small NEOs must be used.

That the average shape for NEOs aligns with the expected elongation for objects formed in collisional interactions suggests that this is indeed how objects in this population were initially formed. The discrepancy between the average shape of NEOs and MBAs is therefore of great interest as it implies that the evolution of these two populations must differ.

The shape of all collisionally formed objects may tend toward the shape described by \cite{michikami2016}. The collisional timescale for NEOs is effectively infinite and hence once these objects enter Near Earth space they will undergo little further shape evolution. In contrast, objects in the main belt will undergo considerable collisional evolution over the timescale of the lifetime of the Solar System. This suggests that collisional evolution drives kilometre-size objects toward more spherical shapes. \cite{henych2015}, however, suggest that collisional evolution should drive objects toward more elongated shapes instead, the opposite of what is seen here. \cite{henych2015} demonstrate that the timescale over which this reshaping occurs for kilometre-size bodies is longer than the collisional disruption timescale of these objects so this is not necessarily a direct contradiction. This invoking of collisional evolution to explain the difference in elongation between MBAs and NEOs is currently purely speculative as it represents the major difference between the evolution of these two populations. Further investigation of how repeated subcatastrophic collisions with small objects will affect the shape of an asteroid must be considered before this conclusion could be verified.

At present this comparison can only be made between larger NEOs and kilometre-size objects in the main asteroid belt. In future when a more complete sample of sub-kilometre MBAs is available it will be possible to determine if the shape discrepancy is also size-dependent, which will give more additional insight as to if the shape discrepancy is indeed collisionally driven.

\section{Conclusions}

We have analysed 867 partial infrared lightcurves for sub-kilometre near Earth objects as observed by {\it Spitzer}. From this sample we have been able to determine an estimate of the shape distribution of NEOs in this size range. The average axis ratio found for this population is $\frac{b}{a} = 0.72 \pm 0.08$. Repeating this using sparse photometry of similarly-sized NEOs observed by Pan-STARRS 1, we find an average axis ratio $\frac{b}{a} = 0.70 \pm 0.10$, showing the shape distributions to be in excellent agreement. The obtained average axis ratio was also in agreement, to within uncertainties, with the value obtained for the main asteroid belt by both \cite{mcneill2016} and \cite{cibulkova2017} ($b/a = 0.8$) also from PS1 survey data. The size range covered by this MBA data set was not the same as that of our NEOs as there is currently no complete data set for sub-kilometre asteroids in the main belt. We constrain a shape distribution for a small sample of kilometre-size NEOs from PS1 data and find that with an average axis ratio $b/a = 0.70 \pm 0.12$ these objects are on average more elongated than their counterparts in the main asteroid belt. The uncertainties on this value are fairly large and a larger data set will be needed to constrain this further. When future surveys such as LSST come online it will be possible to make a more apt comparison between sub-kilometre asteroid populations.

\acknowledgments
\textbf{Acknowledgments}

We thank the anonymous referee for their comments on this manuscript which improved the overall quality of the paper.

The Pan-STARRS1 Surveys (PS1) and the PS1 public science archive have been made possible through contributions by the Institute for Astronomy, the University of Hawaii, the Pan-STARRS Project Office, the Max-Planck Society and its participating institutes, the Max Planck Institute for Astronomy, Heidelberg and the Max Planck Institute for Extraterrestrial Physics, Garching, The Johns Hopkins University, Durham University, the University of Edinburgh, the Queen's University Belfast, the Harvard-Smithsonian Center for Astrophysics, the Las Cumbres Observatory Global Telescope Network Incorporated, the National Central University of Taiwan, the Space Telescope Science Institute, the National Aeronautics and Space Administration under Grant No. NNX08AR22G issued through the Planetary Science Division of the NASA Science Mission Directorate, the National Science Foundation Grant No. AST-1238877, the University of Maryland, Eotvos Lorand University (ELTE), the Los Alamos National Laboratory, and the Gordon and Betty Moore Foundation. We thank the PS1 Builders and PS1 operations staff for construction and operation of the PS1 system and access to the data products provided. 

This work is based on observations made with the {\it Spitzer} Space Telescope, which is operated by the Jet Propulsion Laboratory, California Institute of Technology under a contract with NASA. Support for this work was provided by NASA through an award issued by JPL/Caltech.

This work is supported in part by NSF award 1229776 and NASA award NNX12AG07G. 

%\facility{facility ID}
\facilities{Spitzer (IRAC), Pan-STARRS 1} 
\software{Numpy (\citealt{numpy}), Astropy (\citealt{astropy})}

% \bibliographystyle{yahapj}
% \bibliography{references}

\end{document}